\documentclass[aps,prb,preprint,superscriptaddress,showkeys]{revtex4}
\usepackage{graphicx}
\usepackage{dcolumn}
\usepackage{bm}
\usepackage{amsmath}
\usepackage{amssymb}
\usepackage{latexsym}
\usepackage{epsfig}
\usepackage{amsbsy}
\usepackage{array}
\usepackage{amssymb}
\usepackage{setspace}
\usepackage{bm}
\usepackage{textcomp}
\usepackage{subfig}
\usepackage{epstopdf}

\def\sint{\ifmmode{- \!\!\!\!\!\! \int}
    \else{\hbox{$- \!\!\!\! \int \ $}}\fi}

\begin{document}

\title{The Thermodynamic Transitions of Antiferromagnetic Ising Model on the
Fractional Multi-branched Husimi Recursive Lattice}
\author{Ran Huang \footnote{Correspondence to: ranhuang@sjtu.edu.cn}}
\affiliation{School of Chemistry and Chemical Engineering, Shanghai Jiao Tong University, Shanghai 200240, China}
\author{Chong Chen \footnote{Co-correspondence to: insar.cchen@gmail.com}}
\affiliation{RushandRan Research Co., Shanghai 200433, China}

\begin{abstract}
The multi-branched Husimi recursive lattice has been extended to a virtual structure with fractional numbers of branches joined on one site. Although the lattice is undrawable in real space, the concept is consistent with regular Husimi lattice. The Ising spins of antiferromagnetic interaction on such a sets of lattices were calculated to check the critical temperatures ($T_{c}$) and ideal glass transition temperatures ($T_{k}$) variation with fractional branch numbers. Besides the similar results of two solutions representing the stable state (crystal) and metastable state (supercooled liquid) and indicating the phase transition temperatures, the phase transitions show a well-defined shift with branch number variation. Therefore the fractional branch number as a parameter can be used as an adjusting tool in constructing a recursive lattice model to describe real systems.
\end{abstract}

\maketitle
\section{Introduction}

Husimi lattice as a classical recursive lattice has been intensively studied and employed in many statistical modelings for decades\cite{1,2}. The recursive lattice is believed to be a reliable approximation of regular lattices because of the same coordination numbers. In a recursive lattice the particles fixed on sites have the same interaction environment as in the regular lattices, while the fractal structure avoids the sharing of interactions on neighbor units and consequently provides the advantage of the exact calculation\cite{3,4} Among the works on Husimi lattice, the thermodynamics of Ising model on Husimi is a vigorously interest-drawing subject\cite{5,6,7,8,9,10,11,12,13,14}. Previous works on exact calculation of Ising model on Husimi successfully presented comparable results with other techniques, e.g. mean-field approximation, Monte Carlo or series expansion\cite{15,16,17,18}.

With the nature of the approximation to regular lattices, the setup of the coordination number is critical for the Husimi lattice modeling. There are two ways to settle the coordination number: the basic unit selection and the number of units joined on site, i.e. the number of branches. The original structure of Husimi lattice is featured as recursively constructed by square units (Fig.1a) and it has the coordination number of 4, and the extensions of the same principle with units such as hexagon, tetrahedron and cube, correspondingly with coordination number of 4, 6 and 6 have also been studied and applied in different physical modeling\cite{19,20,21}.

In this work, we follow the second way and the Husimi square lattice is extended to be multi-branched structure. Different than others' works on branch number setup, besides the 2-Dimensional case of 2 branches on one site with coordination number of 4 and 3-Dimensional case of 3 branches with coordination number of 6, the branch number is treated as a variable parameter, which is unnecessary an integer, in the model formulation and calculation. Although the fractional number of branches makes the structure virtual, or to say, cannot be realized in drawing, the calculation shows consistent and realistic results. In section II we will review the lattice geometry of original 2D, and multi-branched 3D case of Husimi lattice. In section III the Ising spins modeled on Husimi lattice will be calculated by recursive calculation technique. In section IV the parameter representing branch number will be set as fractional values and we are going to check the thermodynamics under this circumstance. And Section V is the summary and conclusion.  

\section{Lattice Geometry and Ising model on Husimi lattice}

\begin{figure}
     \centering{
     \subfloat[]{
     \includegraphics[width=0.5\textwidth]{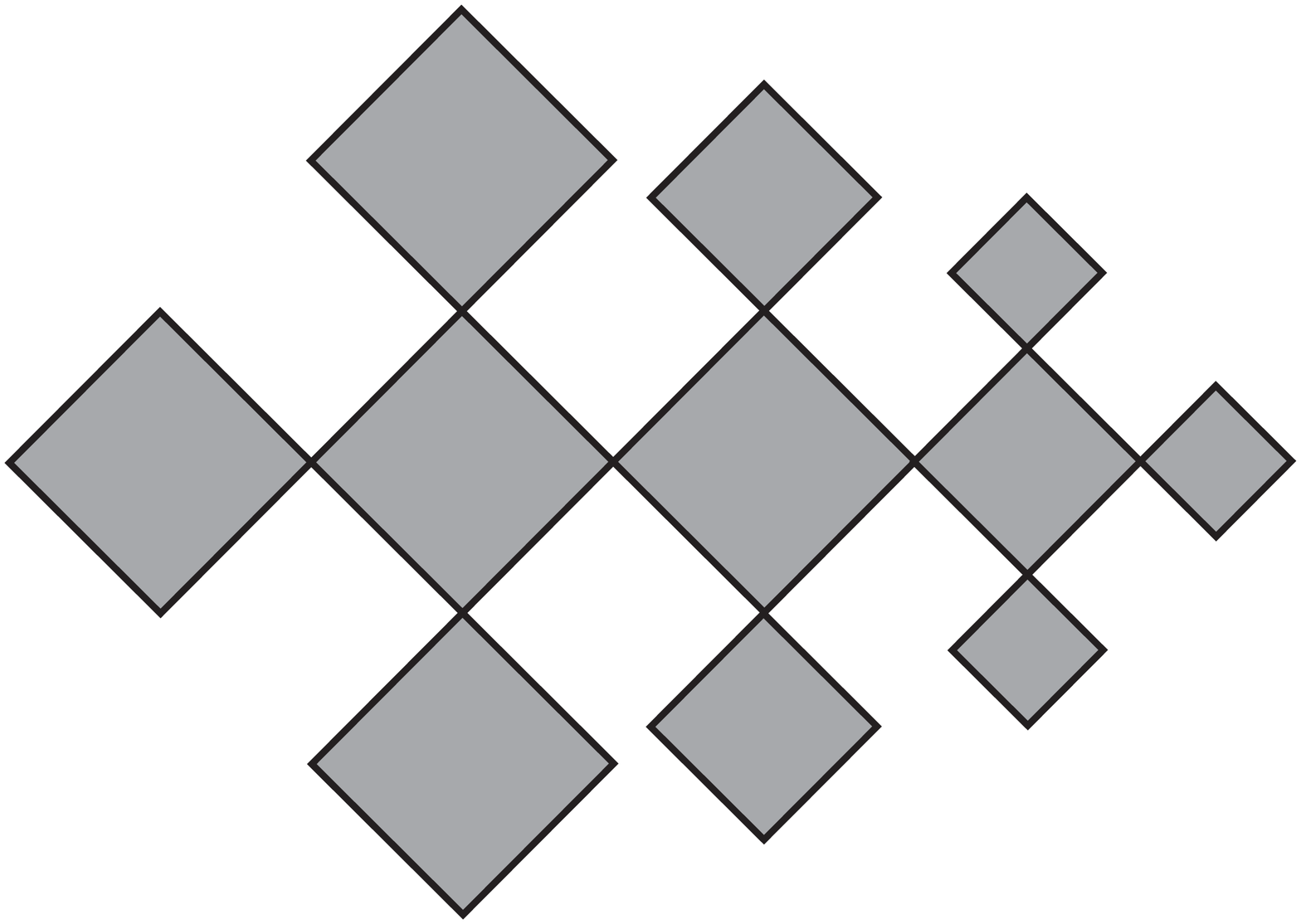}
     }
     \subfloat[]{
     \includegraphics[width=0.4\textwidth]{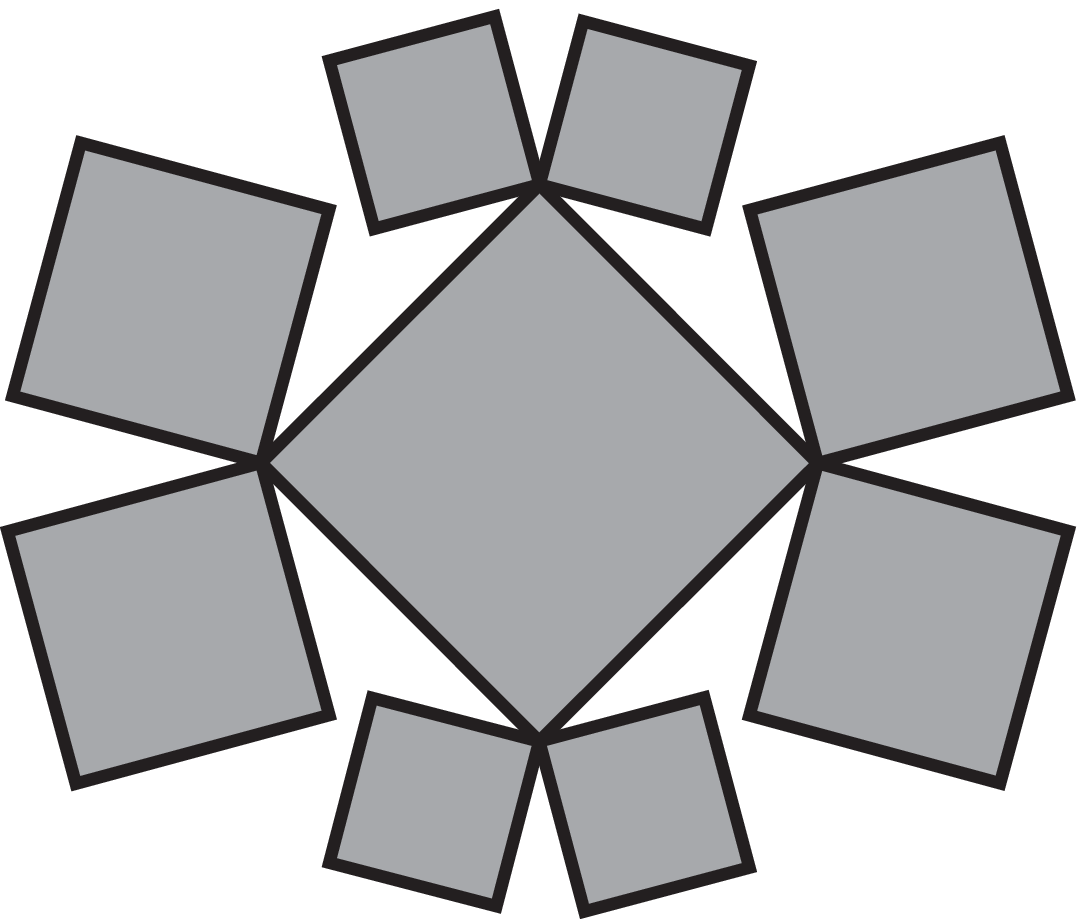}
     }
     }
 \caption{The structure of Husimi square lattice with branches 2 (a) and 3 (b). Note that all squares have the same size and all the bonds represent the same distance or interactions regardless of the length drawn on the page. \label{fig:lattice}}
 \end{figure}

Original Husimi square lattice is featured as a uniformly fractal structure in which one square unit linking other four identical squares on its four sites, it is an approximation to a regular square lattice (the 2-Dimensional case) because the coordination numbers are as the same as four. Extending this recursive construction method can provide a structure with more units linked on one site, for example taking the number of branches $ B=3 $ with the coordination number of six may simulate the 3-Dimensional Case (a regular cubic lattice). Previous works have been examined this 3D representation to be a reliable model and compared it with the cubic recursive representation\cite{18}. Fig.\ref{fig:lattice} shows the structure of Husimi square lattice with branches 2 and 3.

Now we apply Ising spins $S_{i}=\pm1$ on the sites of Husimi lattice. The interaction between two neighbor sites depends on the states of two spins and the interaction parameter $J$, and energy of the entire system is the sum of the interactions:
\begin{equation}
E=\underset{<i,j>}{\sum}-J_{ij}S_{i}S_{j}.
\end{equation}

In this work we do not consider the magnetic field, i.e. the energy of particle itself, and taking a uniform interaction parameter $J$ to be negative to make the spins preferring alternative arrangement for the lowest energy state. In this way the $S=+1$, $-1$, $+1$... alternating state corresponds to the stable (crystal) state, other arrangements with an obtainable stationary solution represent the metastable states.
With four binary spins on the sites one square unit has four interactions and $2^4=16$ possible configurations, the Boltzmann weight of one configuration $\gamma$ is

\begin{equation}
w(\gamma)=exp(-\beta\underset{<i,j>}{\overset{4}{\sum}}-JS_{i}S_{j}),
\end{equation}
where $\beta=1/k_{B}T$, in the following discussion the Boltzmann weight $k_{B}$ is set to be 1 and we set $J=-1$ as the reference temperature scale.
Then the partition function of the entire lattice is

\begin{equation}
Z=\underset{\Gamma}{\sum}{\prod_{\alpha}}w(\gamma_{\alpha}), \label{PF}%
\end{equation}
the $\Gamma={\textstyle\bigotimes_{\alpha}}\gamma_{\alpha}$ denotes the state of the lattice as an ensemble of unit $\alpha$.
Taking the original 2-branched Husimi lattice as an example, for any site in the lattice there are two independent identical sub-trees contributing onto it. Define the partial partition function (PPF) $Z_{i}(S_{i})$ as the contribution of one sub-tree onto the site $S_{i}$, e.g. the $Z_{i}(+1)$ is the sum of weights of one sub-tree with $S_{i}=+1$. Then if we name a random site as the origin point $S_{0}$ of the whole lattice, the partition function can be expressed by the sum of two terms of PPFs contributing to $S_{0}=\pm1$:

\begin{equation}
Z_{0}=Z_{0}^{2}(+1)+Z_{0}^{2}(-1).
\end{equation}

The exponent of 2 is because of two sub-trees (branches) contributing onto $S_{0}$, therefore for a generalized $B$-branched lattice we have the total partition function:
 \begin{equation}
 Z_{0}=Z_{0}^{B}(+1)+Z_{0}^{B}(-1), \label{PF_Origin}
 \end{equation}
and for the lattice shown in Fig.1b we have $B=3$.

\section{Calculations of Ising Model on Husimi Lattice}

Starting from Eq.{\ref{PF_Origin}}, like the total partition function as a function of PPFs, the PPF on level 0 (level $n$ refers to the sub-trees contributing to $S_{n}$) can similarly be expressed as function of lower levels' PPFs plus the local weight. Taking the $n$th level PPF as an example:

 \begin{equation}
Z_{n}(+)=\underset{\gamma=1}{\overset{8}{\sum}}Z_{n+1}^{B^{\prime}%
}(S_{n+1})Z_{n+1}^{B^{\prime}}(S_{n+1}^{\prime})Z_{n+2}^{B^{\prime}}%
(S_{n+2})w(\gamma),\label{PPF+_Recursion}
\end{equation}
\begin{equation}
Z_{n}(-)=\underset{\gamma=9}{\overset{16}{\sum}}Z_{n+1}^{B^{\prime}%
}(S_{n+1})Z_{n+1}^{B^{\prime}}(S_{n+1}^{\prime})Z_{n+2}^{B^{\prime}}%
(S_{n+2})w(\gamma). \label{PPF-_Recursion}
 \end{equation}
By defining $B^{\prime}=B-1$, in Eq.\ref{PPF+_Recursion} and \ref{PPF-_Recursion} on the two sites $S_{n+1}$ and $S_{n+1}^{\prime}$ neighboring to $S_{n}$ there are $B^{\prime}$ sub-trees contributing onto level $n+1$, and similar to the $S_{n+2}$ diagonal to $S_{n}$. The term $w(\gamma)$ is the local weight of the square confined by the four sites $S_{n}$, $S_{n+1}$, $S_{n+1}^{\prime}$, and $S_{n+2}$.

With the PPFs we can introduce a ratio $x(S_{n})$ (shorten as $x_{n}$) on one site%

\begin{equation}
x_{n}=\frac{Z_{n}(+)}{Z_{n}(+)+Z_{n}(-)},\label{Ratios}%
\end{equation}
Since $x_{n}$ is a function of PPFs and vice versa, and from Eq.\ref{PPF+_Recursion} and \ref{PPF-_Recursion} PPFs are also functions of the PPFs on lower two levels, then $x_{n}$ must also be a function of $x$s on lower levels, we then can derive $x_{n}$ by the recursive relation:

\begin{equation}
x_{n}=f(x_{n+1},x_{n+2}). \label{x_recursion}%
\end{equation}
With the solution $x$ and partition functions $Z$ we can exactly calculate the thermodynamics of the system, i.e. the free energy and entropy per spin, and the energy density. The detailed derivation of $x$ and thermodynamics are presented in appendix.

The $x$ is so-called ``solution" of the model and it determines the probability that one site is occupied by the spin $S=+1$, and subsequently determines the entire configuration and thermodynamics of the system. Two stable solutions can be obtained from the calculation: One is a uniform solution regardless of temperature, and without magnetic field this solution is 0.5 everywhere, which is named as one cycle solution; The other solution presents an alternating form of $x_{1},x_{2}$ on two successive levels at low temperature, which we call a 2-cycle solution:

\begin{equation}
x_{1}=f(x_{2},x_{1}) \text{ and } x_{2}=f(x_{1},x_{2}).\label{2-cycle}
\end{equation}

\begin{figure}
      \centering{
      \includegraphics[width=0.75\textwidth]{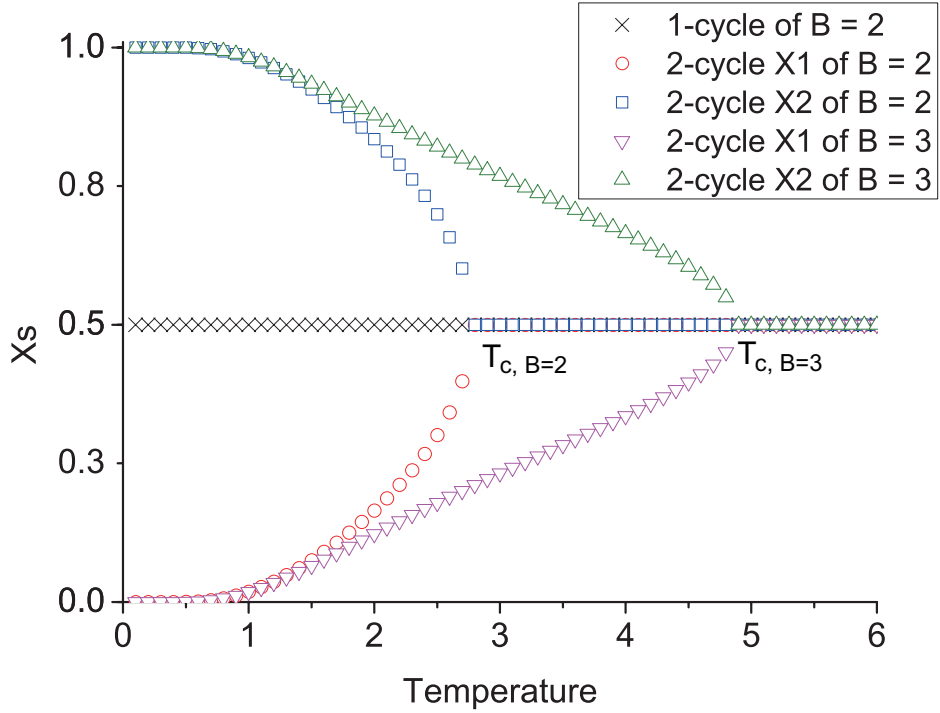}     
      }
 \caption{The 1 and 2-cycle solutions and thermodynamic behaviors of antiferromagnetic Ising model on Husimi square lattice with $B=2$. \label{fig:solution2&3}}
 \end{figure}

The solutions of the 2 and 3-branched Husimi lattice ($B=2$ and $3$) with $J=-1$ is shown in Fig.\ref{fig:solution2&3}. The 1-cycle solution of both systems are $x=0.5$ everywhere, that means the probability of one site occupied by $S=+1$ is fifty percent regardless of temperature, while the 2-cycle solution is also 0.5 at high temperature but below some point it differs to be two solutions, one approaches to $1$ and the other branch approaches to $0$ as $T\rightarrow0$, i.e. we have more probability that $S=+1$ and $S=-1$ alternatively occupying neighboring sites in the lattice, and that probability becomes $100\%$ at absolute zero. Recall the antiferromagnetic case we introduced before, obviously the 2-cycle is lowest-energy preferred and represents the ordered states (crystal) and 1-cycle implies the metastable state (it is still a stable solution but not with the lowest energy). The 2-cycle solution of $B=3$ (the 1-cycle is not shown because it is also 0.5 and therefore not interesting) has a similar behavior with however higher transition temperatures, which can be expected as the more coordination sites brings larger interactions.

\begin{figure}
       \centering{
       \includegraphics[width=0.65\textwidth]{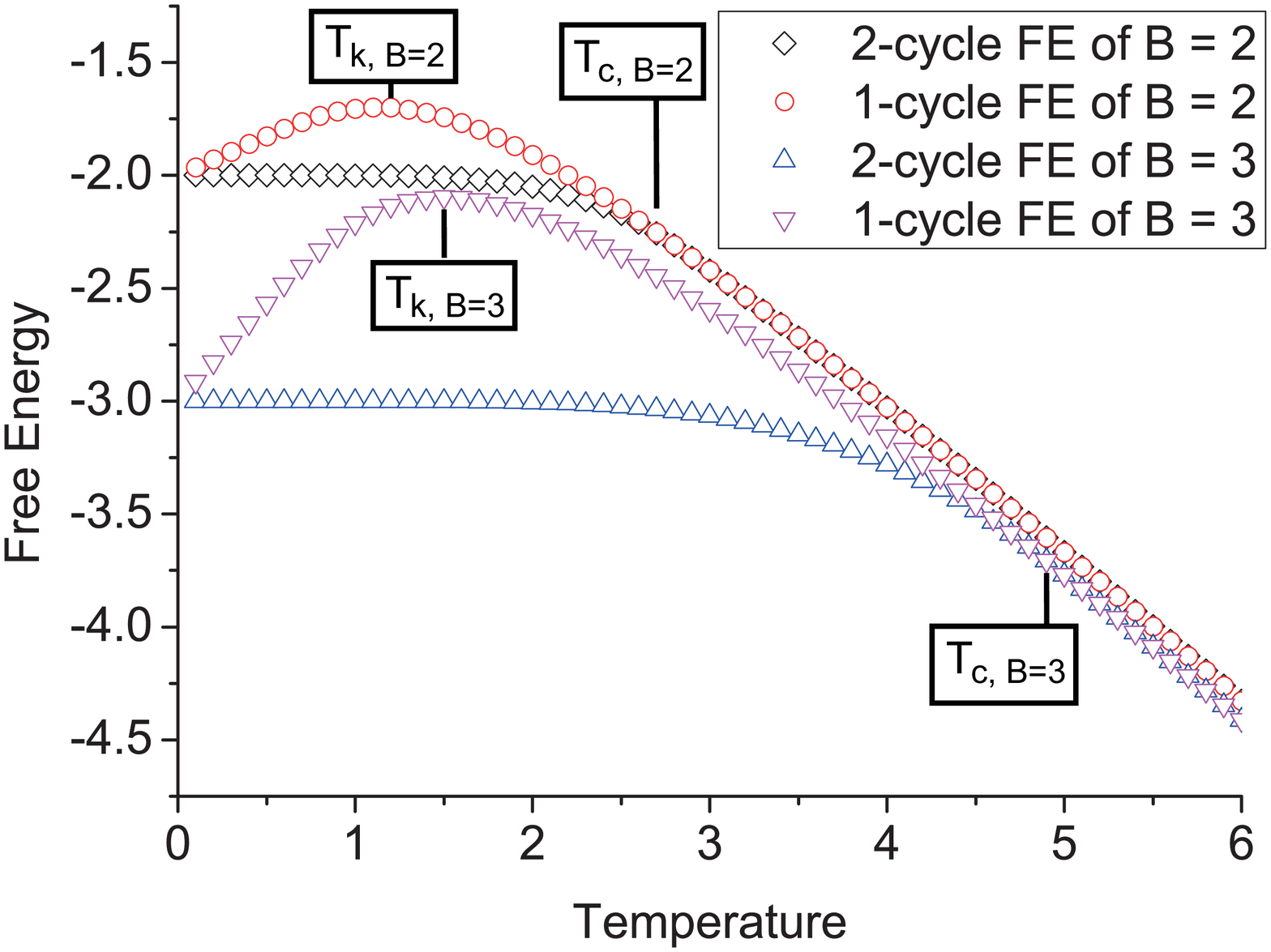} 
       }
  \caption{The 2-cycle solutions and thermodynamic behaviors of antiferromagnetic Ising model on Husimi square lattice with $B=3$. \label{fig:fe2&3}}
  \end{figure}

\begin{figure}
         \centering{
         \includegraphics[width=0.65\textwidth]{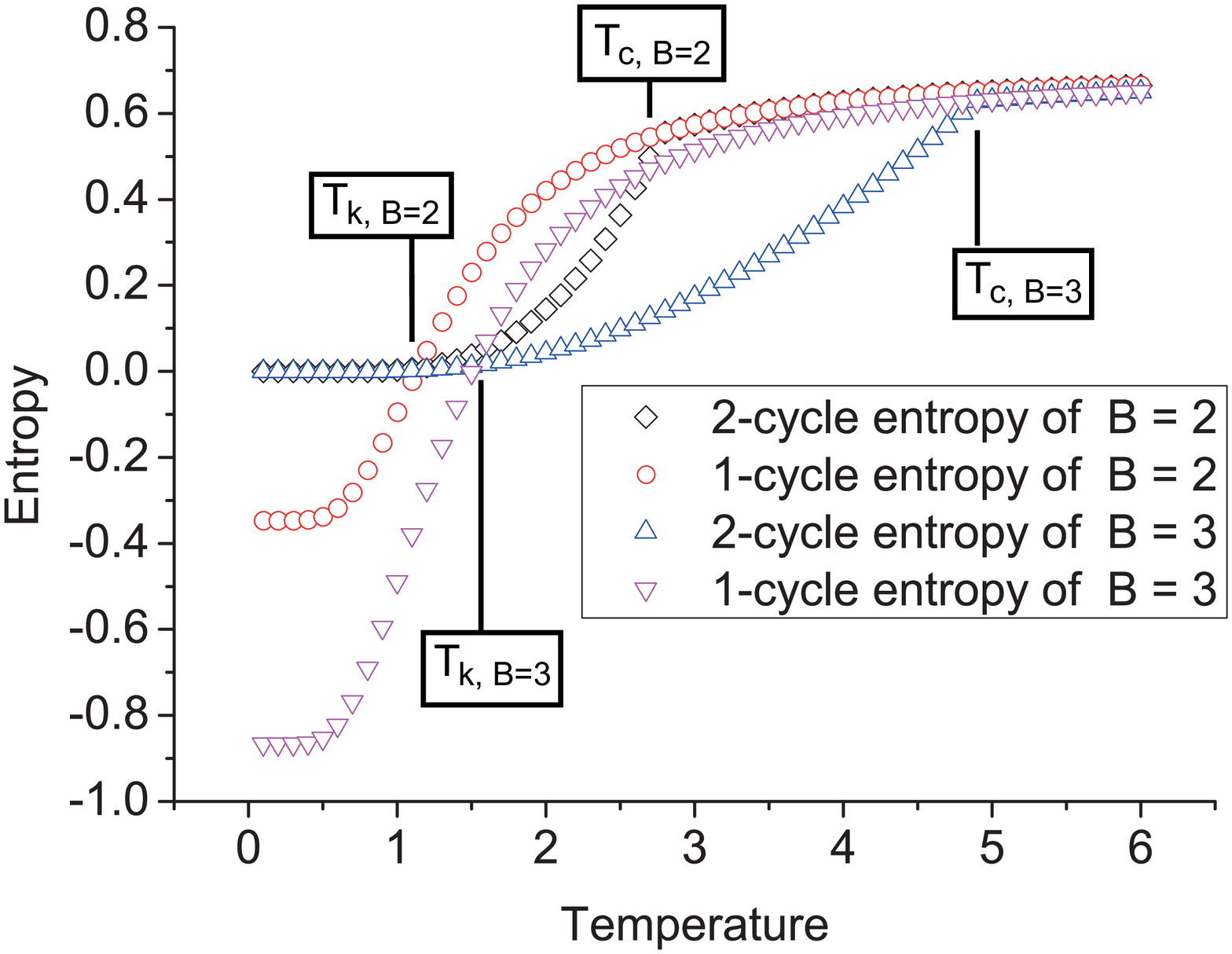}         
         }
    \caption{The 2-cycle solutions and thermodynamic behaviors of antiferromagnetic Ising model on Husimi square lattice with $B=3$. \label{fig:entropy2&3}}
    \end{figure}

Two transitions can be indicated by the thermodynamics of 1 and 2-cycle solutions (Fig.\ref{fig:fe2&3}: free energy and Fig.\ref{fig:entropy2&3}: entropy). With the cooling process at the critical temperature $T_{c}$, where the 2-cycle solution occurs, the system may undergo an ordering transition (crystallization featured as entropy drop in the 2-cycle curves), or stay in the metastable state (supercooled liquid featured as continuous entropy in the 1-cycle curves). Thus obviously it is the melting transition at $T_{c}$. With the continuous temperature decrease, the entropy of the metastable state will extrapolate to zero at another featured temperature $T_{k}$ and then go to negative (corresponding to the unphysical free energy bending down), which is the Kauzmann paradox and ideal glass transition\cite{22}. Similar to the solution implication, the thermodynamic results indicate that both transition temperatures are higher in the $B=3$ system.

\section{Fractional Number of Branches}
\begin{figure}
      \centering{
      \subfloat[]{
      \includegraphics[width=0.5\textwidth]{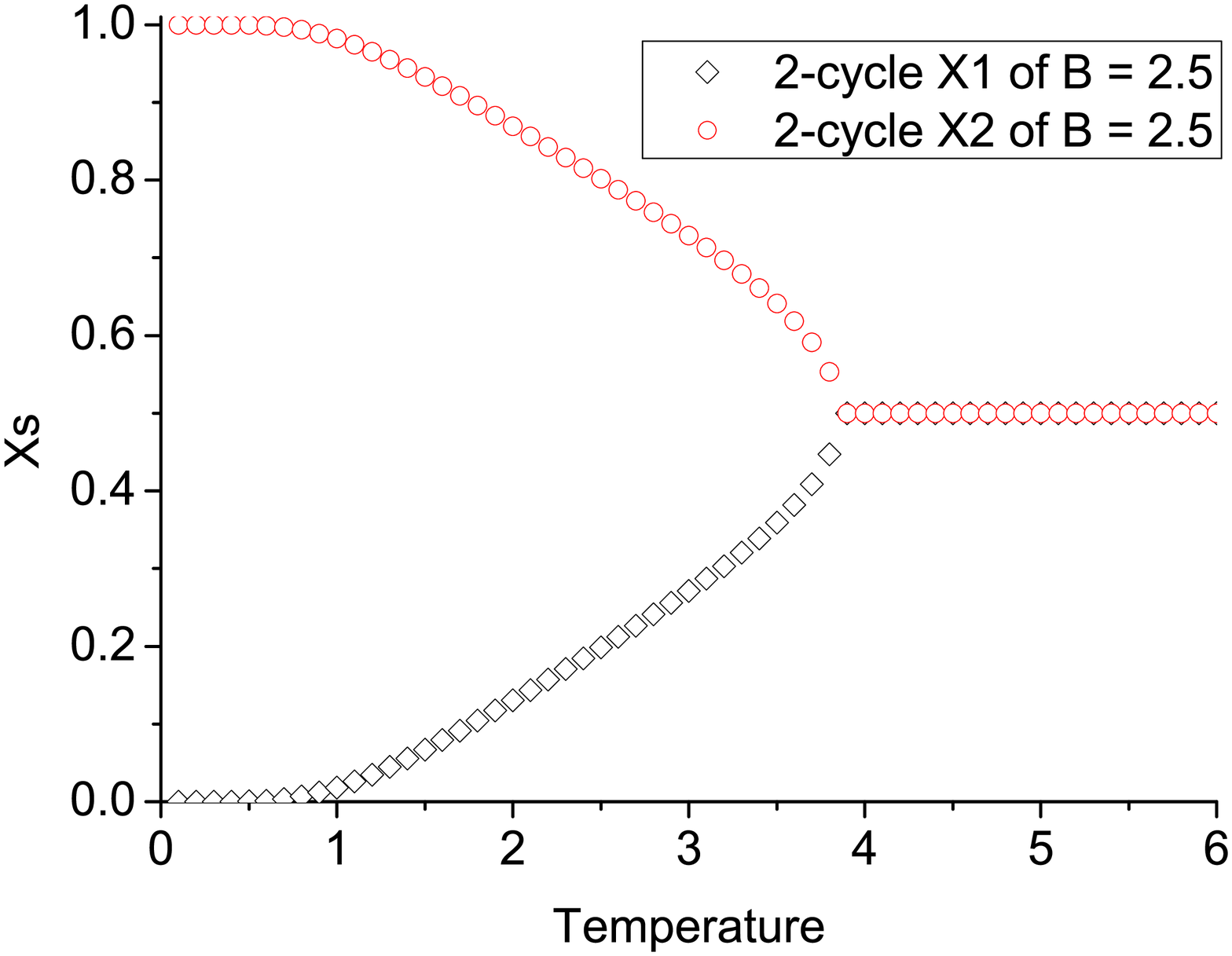} 
      }
      \\
      \subfloat[]{ 
      \includegraphics[width=0.5\textwidth]{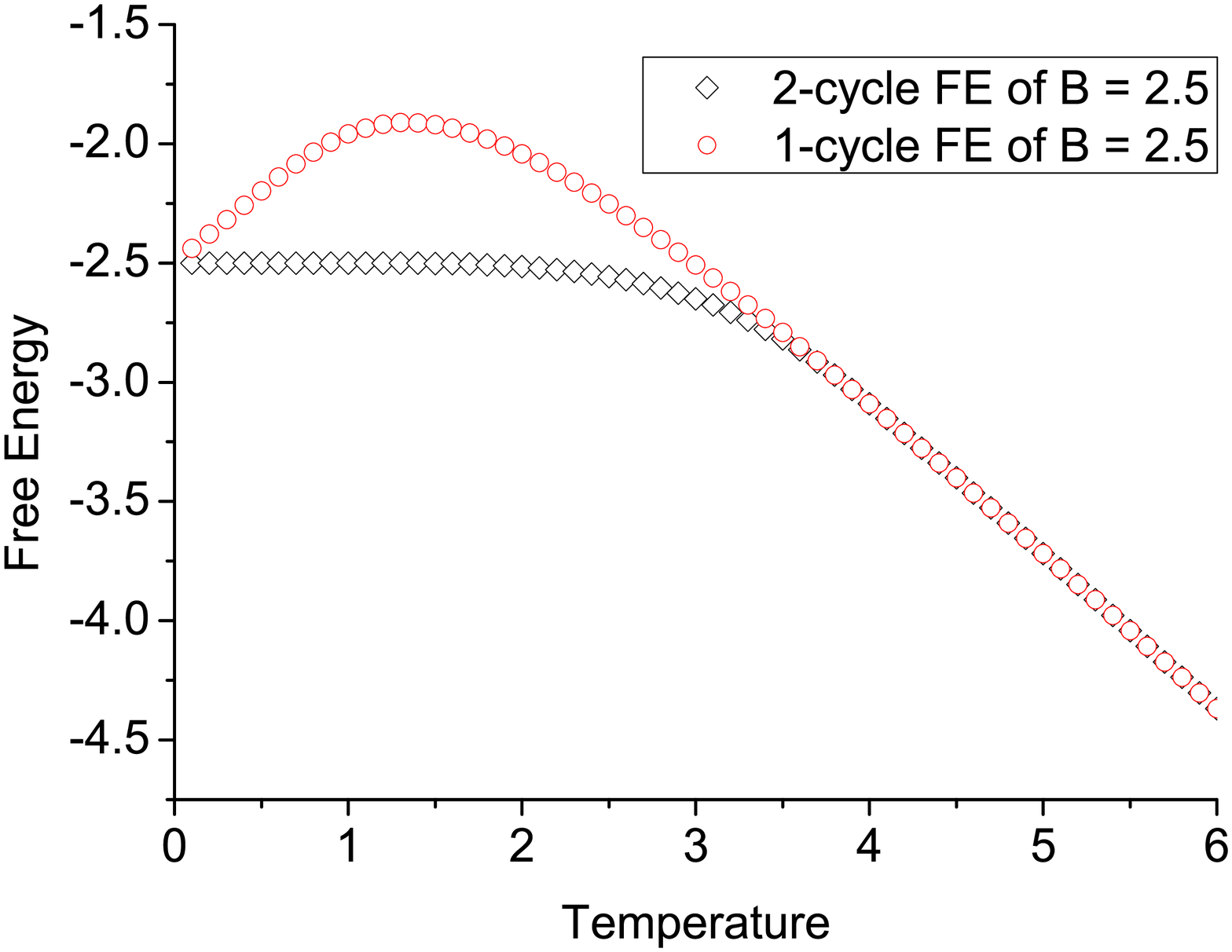}
      }
      \\
      \subfloat[]{ 
      \includegraphics[width=0.5\textwidth]{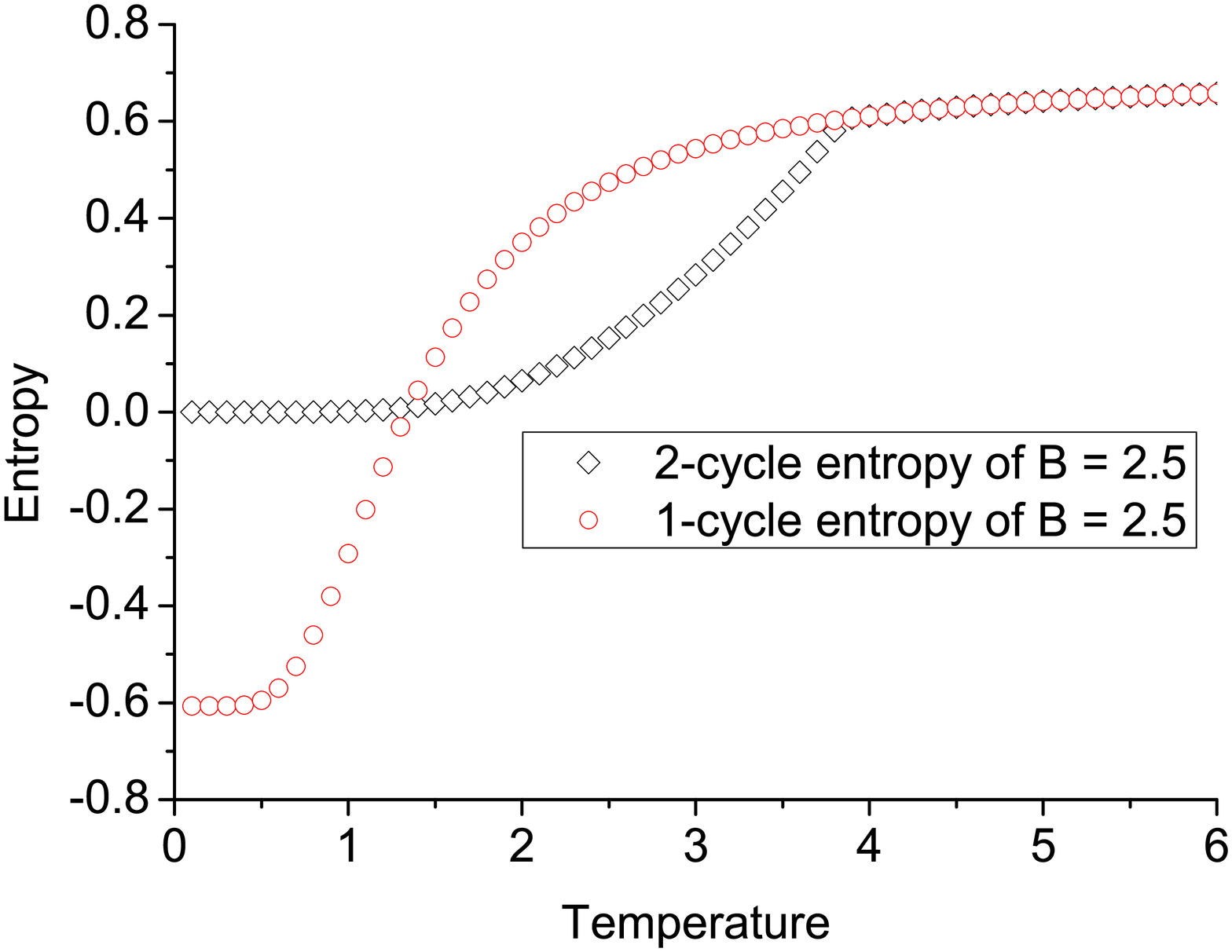}
      }
      }
 \caption{The properties of antiferromagnetic Ising model on Husimi square lattice with $B=2.5$: (a)the 2-cycle solution (b)the free energy (c)the entropy.\label{fig:b=2.5}}
 \end{figure}

\begin{figure}
      \centering{
      \subfloat[]{
      \includegraphics[width=0.7\textwidth]{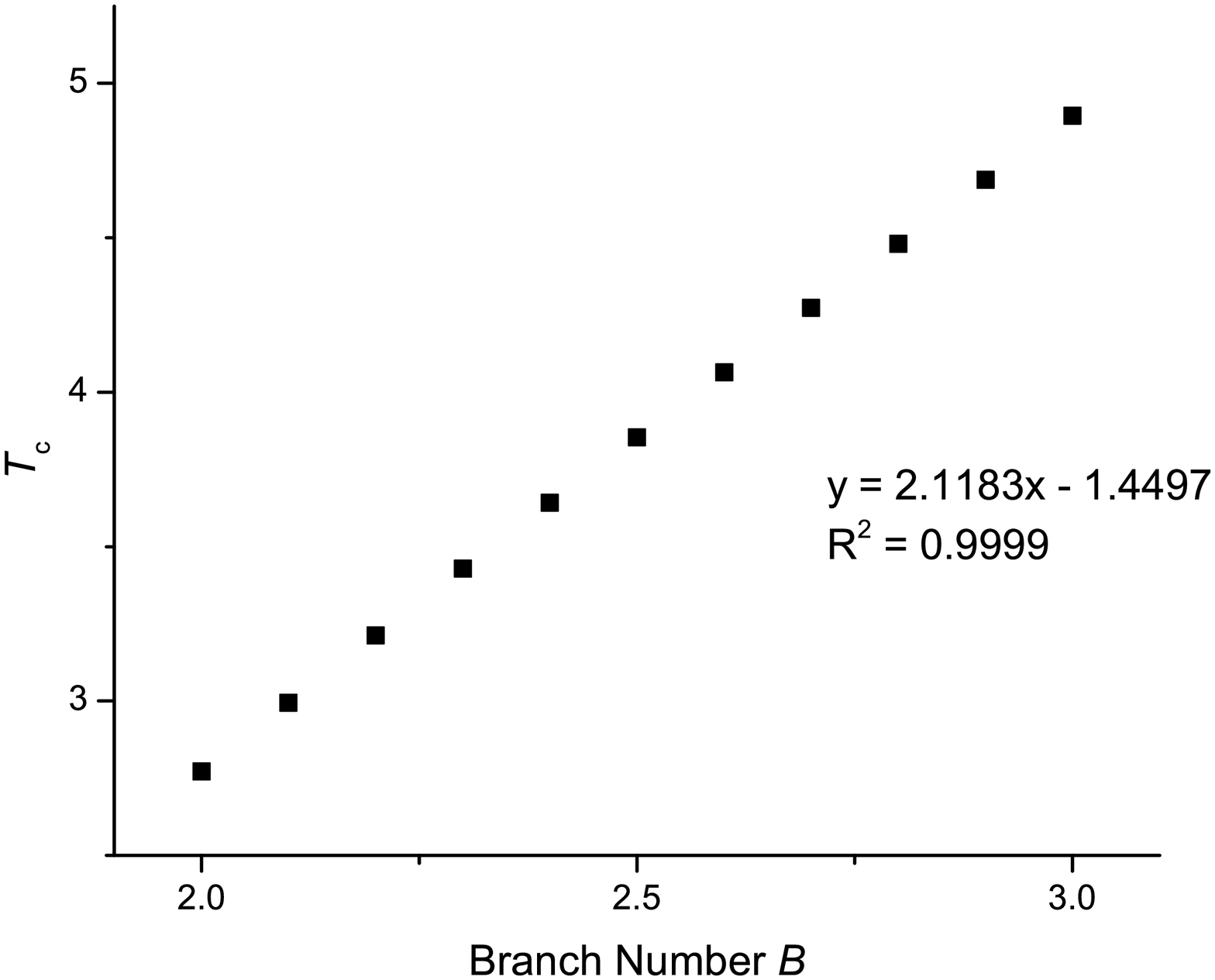}  
      }
      \\
      \subfloat[]{
      \includegraphics[width=0.7\textwidth]{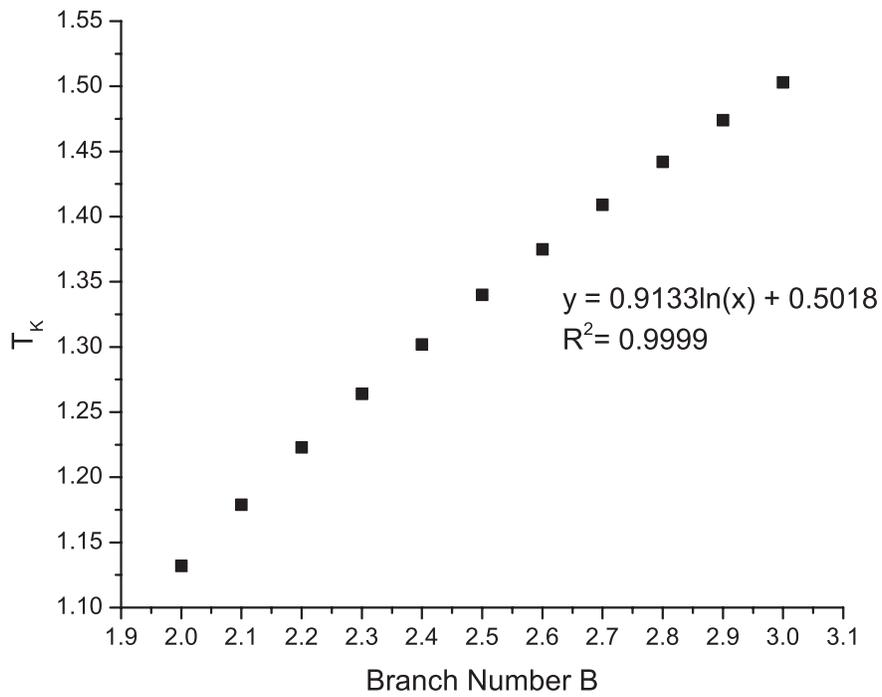}
      }
      }
 \caption{The critical (a) and ideal glass transition temperatures (b) variation with different fractional branch numbers. The $T_{c}$s show a linear behavior and $T_{k}$s has a logarithmic fitting.\label{fig:Tchange}}
 \end{figure}

In the calculation process introduced above, the branch number $B$ is generalized as a parameter in Eq.\ref{PF_Origin}, \ref{PPF+_Recursion} and \ref{PPF-_Recursion}. Without concerning the physical meaning in real space, in the calculation program we are freely to setup $B$ to be fractional values. This setup though does not give any results abnormal, as an example the 2-cycle solution and thermal behaviors of the stable and metastable states with $B=2.5$ are shown in Fig.\ref{fig:b=2.5}. Again the 1-cycle solution is 0.5 and not shown in Fig.\ref{fig:b=2.5}a.
 From Fig.\ref{fig:b=2.5} we can observe the similarly shaped solution and thermodynamics of $B=2.5$: 1) the transitions are in between of these are in $B=2$ and $B=3$; 2) the free energy at near-zero temperature (the ground state energy) is just 2.5, i.e. the coordination number. A series of fractional value of $2<B<3$ have been tested and the transitions temperatures are shown in Fig.\ref{fig:Tchange} and table 1. The $T_{c}$ shows linear increase with $B$, while the $T_{k}$ increase is better to be fitted as a slight logarithmic behavior (very close to linear increase). So far it is not clear about the reason behind the logarithmic fitting of $T_{k}$. However with the observations of as expected, we may conclude that it is safe to take the parameter $B$ as an option control on the transition temperatures and ground state energy, and it provides an tool in fitting the lattice model to describe the real system. With the non-integer value of $B$, the lattice model is generalized to be an abstract concept, however enables it to be more flexible and practical.  

\section{Conclusion}
To summarize, the parameter representing branch numbers in Husimi lattice calculation was set to be non-integer values. This setup virtualized the lattice structure and made it undrawable in real space, however in the program calculation it is merely a simple setting of one parameter. We have checked and confirmed that the imaginary structure can be calculated as a realistic model. The fractional number of branches show nothing abnormal in calculation results for Ising spins modeled on the lattice. In this way, the number of interactions acting on one site directly determined by the coordination number, i.e. the number of branches, which could be any real number in our practice, corresponds to the same value of ground-state energy at near-zero temperature. With subsequently the shift of transition temperatures $T_{c}$ and $T_{k}$, the general thermodynamic behaviors of Ising systems on recursive lattice are consistent, and that implies the branch number may serve as an adjustable option to make the model more flexible in specific fittings to describe real system.

\newpage

\begin{table}[htbp]
  \caption{The transition temperature variations with different fractional branch numbers.}
  \label{tab:table1}
  \begin{center}
  \begin{tabular}{llll}
  \hline\hline
   $B$ & $T_{\text{c}}$ & $T_{\text{k}}$ & $T_{\text{c}}/T_{\text{k}}$\\
   \hline
   2.0 & 2.772 & 1.132 & 2.449\\
   2.1 & 2.995 & 1.179 & 2.540\\
   2.2 & 3.213 & 1.223 & 2.627\\
   2.3 & 3.429 & 1.264 & 2.713\\
   2.4 & 3.642 & 1.302 & 2.797\\
   2.5 & 3.854 & 1.340 & 2.876\\
   2.6 & 4.064 & 1.375 & 2.956\\
   2.7 & 4.273 & 1.409 & 3.033\\
   2.8 & 4.481 & 1.442 & 3.107\\
   2.9 & 4.688 & 1.474 & 3.180\\
   3.0 & 4.895 & 1.503 & 3.257\\
   \hline\hline
  \end{tabular}
  \end{center}
\end{table}

\newpage

\appendix{Appendix}

\subsection{\textbf{The calculation of} $\boldmath{x}$}
By starting from
\begin{align*}
x_{n}=\frac{Z_{n}(+)}{Z_{n}(+)+Z_{n}(-)},\\y_{n}=\frac{Z_{n}(-)}{Z_{n}%
(+)+Z_{n}(-)},
\end{align*}
we define a compact note
\begin{align*}
z_{n}(S_{n})=\left\{
\begin{array}
[c]{c}%
x_{n}\text{ if }S_{n}=+1\\
y_{n}\text{ if }S_{n}=-1
\end{array}
\right.
\end{align*}
In terms of%
\[
A_{n}^{B\prime}=Z_{n}(+)+Z_{n}(-),
\]
we have
\begin{align*}
A_{n}^{B\prime}z_{n}(\pm)  =\sum A_{n+1}^{B\prime}z_{n+1}^{B^{\prime}%
}(S_{n+1})A_{n+1}^{B\prime}z_{m+1}^{B^{\prime}}(S_{n+1}^{\prime}%
)A_{n+2}^{B\prime}z_{n+2}^{B^{\prime}}(S_{n+2})w(\gamma),\\
z_{n}(\pm) =\sum z_{n+1}^{B^{\prime}}(S_{n+1})z_{n+1}^{B^{\prime}}%
(S_{n+1}^{\prime})z_{n+2}^{B^{\prime}}(S_{n+2})w(\gamma)/Q(x_{n+1},x_{n+2}),
\end{align*}
where the sum is over $\gamma=1,2,3,\ldots,8$ for $S_{n}=+1$, and over
$\gamma=9,10,11,\ldots,16$ for $S_{n}=-1$, and where
\[
Q(x_{n+1},x_{n+2})\equiv\left[  A_{n}/A_{n+1}^{2}A_{n+2}\right]^{B^{\prime}%
};
\]
it is related to the polynomials
\begin{align*}
Q_{+}(x_{n+1},x_{n+2}) =\underset{\gamma=1}{\overset{8}{\sum}}%
z_{n+1}^{B^{\prime}}(S_{n+1})z_{n+1}^{B^{\prime}}(S_{n+1}^{\prime}%
)z_{n+2}^{B^{\prime}}(S_{n+2})w(\gamma),\\
Q_{-}(x_{n+1},x_{n+2}) =\underset{\gamma=9}{\overset{16}{\sum}}%
z_{n+1}^{B^{\prime}}(S_{n+1})z_{n+1}^{B^{\prime}}(S_{n+1}^{\prime}%
)z_{n+2}^{B^{\prime}}(S_{n+2})w(\gamma),
\end{align*}
according to
\[
Q(x_{n+1},x_{n+2})=Q_{+}(x_{n+1},x_{n+2})+Q_{-}(y_{n+1},y_{n+2}).
\]
In terms of the above polynomials, we can express the recursive relation for
the ratio $x_{n}$ in terms of $x_{n+1}$ and $x_{n+2}$:
\begin{align*}
x_{n}=\frac{Q_{+}(x_{n+1},x_{n+2})}{Q(x_{n+1},x_{n+2})}.
\end{align*}
Then the 2-cycle solution can be calculated as:
\[
x_{1}=\frac{Q_{+}(x_{2},x_{1})}{Q(x_{2},x_{1})},x_{2}=\frac{Q_{+}(x_{1}%
,x_{2})}{Q(x_{1},x_{2})}.
\]
and the 1-cycle solution is just a special case when $x_{1}=x_{2}$.\\

\subsection{\textbf{Free Energy}}

Since the lattice is infinite, we are only interested in the free energy per site. The calculation technique is following the Gujrati trick in reference\cite{15,16}. Here we give a brief introduction: Consider the $B$ squares meeting at the origin site $S_{0}$. Surrounding $S_{0}$ we have $S_{1}$ and $S_{2}$ on next levels. If we cut off the $BB^{\prime}$ branches on level 1 and hook up them together to form $B^{\prime}$ smaller lattices, the
partition function of each of these smaller lattices is:%
\[
Z_{1}=Z_{1}^{B}(+)+Z_{1}^{B}(-).
\]
Similarly for the cutting off and hook-up of branches on level 2:
\[
Z_{2}=Z_{2}^{B}(+)+Z_{2}^{B}(-).
\]
The free energy of the left out squares is
\[
F_{local}=-T\log\left[  \frac{Z_{0}}{(Z_{1}^{2}Z_{2})^{B^{\prime}}}\right]  .
\]
We have $4/B$ cites in a square and $B$ squares in the local origin region. The free energy per site is:%
\begin{align*}
F=-\frac{F_{local}}{4}. \label{FreeEnergy/site}%
\end{align*}
By substituting $Z_{n}(+)=A_{n}x_{n}$ and $Z_{n}(-)=A_{n}y_{n},$ we have
\begin{align*}
F=-\frac{1}{4}T\log(Q^{2B^{\prime}}\frac{1}{\{[x_{1}^{B}
+(1-x_{1})^{B}]^{2}[x_{2}^{B}+(1-x_{2})^{B}]\}^{B^{\prime}}}).
\end{align*}
Then subsequently the entropy and energy per site can be calculated by the free energy.
\end{document}